\documentclass[11pt]{article}

\newcommand{\beq}{\begin{eqnarray}}
\newcommand{\eeq}{\end{eqnarray}}

\newcommand{\vtd}{|V_{td}|}

\newcommand{\mw}{M_{\rm W}}

\newcommand{\gev}{\, {\rm GeV}}

\newcommand{\bea}{\begin{eqnarray}}
\newcommand{\eea}{\end{eqnarray}}
\newcommand{\bd}{\begin{displaymath}}
\newcommand{\ed}{\end{displaymath}}

\newcommand{\be}{\begin{equation}}
\newcommand{\ee}{\end{equation}}
\newcommand{\bi}{\begin{itemize}}
\newcommand{\ei}{\end{itemize}}
\newcommand{\ord}{{\cal O}}

\renewcommand{\baselinestretch}{1.3}

\begin{document}
\thispagestyle{empty}
\phantom{xxx}
\vskip1truecm
\begin{flushright}
 TUM-HEP-502/03 \\
March 2003
\end{flushright}
\vskip0.8truecm

\begin{center}
 {\Large\bf Relations between \boldmath{$\Delta M_{s,d}$} and
\boldmath{$B_{s,d}\to \mu\bar\mu$}\\ 
in Models with Minimal Flavour Violation}
\end{center}

\vskip0.8truecm
\centerline{\large\bf Andrzej J. Buras}
\bigskip
\centerline{\sl Technische Universit{\"a}t M{\"u}nchen}
\centerline{\sl Physik Department} 
\centerline{\sl D-85748 Garching, Germany}
\vskip1truecm
\centerline{\bf Abstract}
\setcounter{equation}{0}
The predictions for the
$B_{s,d}-\bar B_{s,d}$ mixing mass differences $\Delta M_{s,d}$ and 
the branching ratios $Br(B_{s,d}\to\mu\bar\mu)$ 
within the Standard Model (SM) and its extensions suffer from considerable
hadronic uncertainties present in the $B_{s,d}$-meson 
decay constants $F_{B_{s,d}}$ that enter these quantities quadratically.
We point out that in the restricted class of 
 models with minimal flavour violation (MFV) in which only the SM low 
energy operators are relevant, the ratios 
$Br(B_{q}\to\mu\bar\mu)/\Delta M_q$ $(q=s,d)$
do not depend on $F_{B_{q}}$ and the CKM matrix elements. They involve
in addition to the short distance functions and B meson lifetimes
only the non-perturbative parameters $\hat B_{s,d}$. The latter are under much 
better control than $F_{B_{s,d}}$. 
Consequently in these models the predictions for $Br(B_{q}\to\mu\bar\mu)$
have only small hadronic uncertainties once $\Delta M_q$ are experimentally
known. Of particular interest is also the relation
\bea
\frac{Br(B_{s}\to\mu\bar\mu)}{Br(B_{d}\to\mu\bar\mu)}
=\frac{\hat B_{d}}{\hat B_{s}}
\frac{\tau( B_{s})}{\tau( B_{d})} 
\frac{\Delta M_{s}}{\Delta M_{d}} \nonumber 
\eea
that is practically free of theoretical uncertainties as 
$\hat B_{s}/\hat B_{d}=1$ up to small $SU(3)$ breaking corrections.
Using these ideas within the SM we find much more accurate predictions
than those found in the literature: 
$Br(B_{s}\to\mu\bar\mu)=(3.4\pm 0.5)\cdot 10^{-9}$ and
$Br(B_{d}\to\mu\bar\mu)=(1.00\pm 0.14)\cdot 10^{-10}$ were 
in the first case we assumed as an example $\Delta M_s=(18.0\pm 0.5)$/ps.


\newpage

{\bf 1.} Among the possible extentions of the Standard Model (SM), 
of particular 
interest are the models with {\it minimal flavour violation} (MFV), where 
the only source for flavour mixing is still given by the CKM matrix 
(see, for instance \cite{UUT,DAGIISST,BOEWKRUR}). 
In the restricted class of these models \cite{UUT}, in which 
only the SM low energy operators are relevant, it is possible to derive 
relations between various observables that are independent of the parameters 
specific to a given MFV model \cite{UUT,REL}. Violation of these relations 
would indicate the relevance of new low energy operators and/or the presence 
of new sources of flavour violation encountered for instance in general 
supersymmetric models \cite{Masiero,MIPORO,GRNIRA,DEPI}.

In this letter we would like to point out the existence of simple relations 
between 
the $B_{s,d}-\bar B_{s,d}$ mixing mass differences $\Delta M_{s,d}$ 
and the branching ratios for the rare decays $B_{s,d}\to\mu\bar\mu$ 
that are valid in models with minimal flavour violation (MFV) as defined in
\cite{UUT}. These relations should be of interest for the Run II at Tevatron 
and later for the LHC and BTeV experiments where $\Delta M_s$ and 
$Br(B_{s,d}\to\mu\bar\mu)$ should be measured. Moreover, they allow one 
to make much more accurate predictions for $Br(B_{s,d}\to\mu\bar\mu)$ once
$\Delta M_{s,d}$ are precisely known.
To our knowledge the relations in question have not been discussed so far 
in the literature except for a short comment made by us in \cite{AJB95}.

{\bf 2.} Within the MFV models $\Delta M_{s,d}$ and $Br(B_{s,d}\to\mu\bar\mu)$
are given as follows ($q=d,s$) \cite{BBL}
\begin{equation}\label{DMQ}
\Delta M_q = \frac{G_{\rm F}^2}{6 \pi^2} \eta_B m_{B_q} 
(\hat B_{q} F_{B_q}^2 ) \mw^2  
|V^\ast_{tb}V_{tq}|^2 S(x_t,x_{\rm new}),
\end{equation}
\begin{equation}\label{bbll}
Br(B_q\to \mu\bar\mu)=\tau(B_q)\frac{G^2_{\rm F}}{\pi}\eta_Y^2
\left(\frac{\alpha}{4\pi\sin^2\theta_{W}}\right)^2 F^2_{B_q}m^2_\mu m_{B_q}
|V^\ast_{tb}V_{tq}|^2 Y^2(x_t,\bar x_{\rm new}),
\end{equation}
where $F_{B_q}$ is the $B_q$-meson decay constant and
$\hat B_q$ the renormalization group invariant parameter related 
to the hadronic matrix element of the operator $Q(\Delta B=2)$. See 
\cite{BBL} for details. $\eta_B=0.55\pm0.01$ \cite{BJW90,UKJS} and 
$\eta_Y=1.012$ \cite{BUBU} are the short distance 
QCD corrections evaluated using $m_t\equiv\overline{m}_t(m_t)$. 
In writing (\ref{bbll}) we have neglected the
terms $\ord(m_\mu^2/m_{B_q}^2)$ in the phase space factor. The short 
distance functions $S(x_t,x_{\rm new})$
and $Y(x_t,\bar x_{\rm new})$ result from the relevant box and penguin 
diagrams specific to a given MFV model. They depend on the top quark 
mass ($x_t=m_t^2/M^2_W$) and new parameters like the masses of new particles 
that we denoted collectively by $x_{\rm new}$ and $\bar x_{\rm new}$. 
Explicit expressions
for these functions in the MSSM at low $\tan\beta$ and in the ACD 
model \cite{appelquist:01} in five dimensions can be found in 
\cite{BRMSSM} and \cite{BSW02}, respectively.

The main theoretical uncertainties in (\ref{DMQ}) and (\ref{bbll})
originate in the values of $\hat B_{q} F_{B_q}^2$ and $F_{B_q}^2$ 
for which the most recent published values obtained by lattice simulations 
read  \cite{lellouch}
\be\label{input1}
F_{B_d}\sqrt{\hat B_{d}}=(235\pm33^{+0}_{-24})~{\rm MeV}, \qquad 
F_{B_s}\sqrt{\hat B_{s}}=(276\pm 38)~{\rm MeV},
\ee
\be\label{input2}
F_{B_d}=(203\pm27^{+0}_{-20})~{\rm MeV}, \qquad 
F_{B_s}=(238\pm 31)~{\rm MeV}~.
\ee
Similar results are obtained by means of QCD sum rules \cite{Jamin}.
Consequently the hadronic uncertainties in
$\Delta M_{s,d}$ and $Br(B_{s,d}\to\mu\bar\mu)$ are in the ballpark
of $\pm 30\%$ which is clearly disturbing.
The uncertainties in the $B_q$-meson lifetimes are substantially 
smaller \cite{ref:lepblife}:  
\be\label{LIFE}
\tau(B_s)=(1.461\pm 0.057)~{\rm ps},\qquad 
\tau(B_d)=(1.540\pm 0.014)~{\rm ps}, \qquad
\frac{\tau(B_s)}{\tau(B_d)}=0.949\pm 0.038~.
\ee
As noticed by many authors in the past, the uncertainties in 
$\Delta M_{s,d}$ and  $Br(B_{s,d}\to\mu\bar\mu)$
can be considerably reduced by considering the ratios
\be\label{dmsdmd}
\frac{\Delta M_d}{\Delta M_s}=
\frac{m_{B_d}}{m_{B_s}}
\frac{\hat B_{d}}{\hat B_{s}}\frac{F^2_{B_d}}{F^2_{B_s}}
\left|\frac{V_{td}}{V_{ts}}\right|^2
\end{equation}
\begin{equation}\label{bmumu}
\frac{Br(B_d\to\mu^+\mu^-)}{Br(B_s\to\mu^+\mu^-)}=
\frac{\tau({B_d})}{\tau({B_s})}\frac{m_{B_d}}{m_{B_s}}
\frac{F^2_{B_d}}{F^2_{B_s}}
\left|\frac{V_{td}}{V_{ts}}\right|^2
\end{equation}
that can be used to determine $|V_{td}/V_{ts}|$ without the pollution
from new physics \cite{UUT}. In particular the relation (\ref{dmsdmd}) 
will offer
after the measurement of $\Delta M_s$ a  powerful determination of the 
length of one side of the unitarity triangle, denoted usually by $R_t$.
As 
\cite{lellouch}
\be
\xi=\frac{\sqrt{\hat B_{s}} F_{B_s}}{\sqrt{\hat B_{d}}F_{B_d}}\approx
\frac{F_{B_s}}{F_{B_d}}=
1.18\pm 0.04^{+0.12}_{-0}~,
\ee
(see also $\xi=1.22\pm 0.07$ \cite{BEFAPRZU}) 
the uncertainties in the relations (\ref{dmsdmd}) and (\ref{bmumu}) 
are in the ballpark of $\pm 15\%$ and thus by roughly a factor of two
smaller than in (\ref{DMQ}) and (\ref{bbll}).

{\bf 3.} Here we would like to point out three useful relations that do not 
involve the decay constants $F_{B_d}$ and consequently contain 
substantially smaller hadronic uncertainties than the formulae considered 
so far. These relations follow directly from (\ref{DMQ}) and (\ref{bbll}) 
and read
\be\label{R1}
\frac{Br(B_{s}\to\mu\bar\mu)}{Br(B_{d}\to\mu\bar\mu)}
=\frac{\hat B_{d}}{\hat B_{s}}
\frac{\tau( B_{s})}{\tau( B_{d})} 
\frac{\Delta M_{s}}{\Delta M_{d}},
\ee
\be\label{R2}
Br(B_{q}\to\mu\bar\mu)
=C\frac{\tau(B_{q})}{\hat B_{q}}
\frac{Y^2(x_t,\bar x_{\rm new})}{S(x_t,x_{\rm new})} 
\Delta M_{q}, \qquad (q=s,d)
\ee
with 
\be
C=\eta_Y^2\frac{6\pi}{\eta_B}
\left(\frac{\alpha}{4\pi\sin^2\theta_{W}}\right)^2\frac{m^2_\mu}{\mw^2}
=4.36\cdot 10^{-10}
\ee
where we have used $\alpha=1/129$, $\sin^2\theta_{W}=0.23$ and 
$\mw=80.423\gev$ \cite{PDG}.

The relevant parameters obtained from lattice simulations are
\cite{lellouch}
\be\label{BBB}
\frac{\hat B_{s}}{\hat B_{d}}=1.00\pm 0.03, \qquad
\hat B_{d}=1.34\pm0.12, \qquad \hat B_{s}=1.34\pm0.12~.
\ee
The simple relation between $\Delta M_s/\Delta M_d$ and 
$Br(B_{s}\to\mu\bar\mu)/Br(B_{d}\to\mu\bar\mu)$ in (\ref{R1}) involves
only measurable quantities except for the ratio $\hat B_{s}/\hat B_{d}$
that has to be calculated by non-perturbative methods.
As $\hat B_{s}/\hat B_{d}=1$ in the SU(3) flavour symmetry limit, only 
SU(3) breaking corrections have to be calculated. 
Now, in contrast to $F_{B_s}/F_{B_d}$ and $F_{B_d}$ that suffer from 
chiral logarithms and quenching \cite{KRRY,lellouch,BEFAPRZU}, the chiral
extrapolation in the case of $\hat B_{q}$ is well controlled and very 
little variation is observed between quenched and $N_f=2$ results.  
Consequently the error in $\hat B_{s}/\hat B_{d}=1$ is very small and 
also the separate values for $\hat B_{s}$ and $\hat B_{d}$ 
given in (\ref{BBB}) are rather accurate 
\cite{KRRY,lellouch,BEFAPRZU,Yamada}.
These results should be further improved in the future.
Consequently (\ref{R1}) is one of the cleanest relations in B physics but
also (\ref{R2}) is rather clean theoretically.

We note that once $\Delta M_s/\Delta M_d$ has been precisely measured, 
the relation (\ref{R1}) will allow one to predict 
$Br(B_{s}\to\mu\bar\mu)/Br(B_{d}\to\mu\bar\mu)$ without essentially 
any hadronic uncertainties. On the other hand the relations in (\ref{R2}) 
allow to predict $Br(B_{s,d}\to\mu\bar\mu)$  
in a given MFV model with substantially smaller hadronic uncertainties 
than found by using directly the formulae in (\ref{bbll}). In particular using
the known formulae for the functions $Y$ and $S$ in the SM model
\cite{BBL}, we find
\be\label{R3}
Br(B_{s}\to\mu\bar\mu)
=3.42\cdot 10^{-9}\left[\frac{\tau(B_{s})}{1.46~ {\rm ps}}\right]
\left[\frac{1.34}{\hat B_{s}}\right]
\left[\frac{\overline{m}_t(m_t)}{167\gev}\right]^{1.6}
\left[\frac{\Delta M_{s}}{18.0/{\rm ps}}\right], 
\ee

\be\label{R4}
Br(B_{d}\to\mu\bar\mu)
=1.00\cdot 10^{-10}\left[\frac{\tau(B_{d})}{1.54 ~{\rm ps}}\right]
\left[\frac{1.34}{\hat B_{d}}\right]
\left[\frac{\overline{m}_t(m_t)}{167\gev}\right]^{1.6}
\left[\frac{\Delta M_{d}}{0.50/{\rm ps}}\right]. 
\ee

Using $\overline{m}_t(m_t)=(167\pm 5)\gev$, the lifetimes in (\ref{LIFE}), 
$\hat B_q$ in (\ref{BBB}), $\Delta M_d=(0.503\pm0.006)/{\rm ps}$ 
\cite{ref:lepbosc}
 and taking as an example 
$\Delta M_s=(18.0\pm0.5)/{\rm ps}$ we find the predictions for the branching
ratios in question
\be\label{Results}
 Br(B_{s}\to\mu\bar\mu)=(3.42\pm 0.54)\cdot 10^{-9}, \qquad
Br(B_{d}\to\mu\bar\mu)=(1.00\pm 0.14)\cdot 10^{-10}.
\ee
These results are substantially more accurate than the ones found in the 
literature (see for instance \cite{DAGIISST,BOEWKRUR,Erice,FIM03}) where
the errors are in the ballpark of $\pm (30-50)\%$.

In calculating the errors in (\ref{Results}) we have added first the 
experimental errors in $\tau(B_{q})$, $\overline{m}_t(m_t)$ and 
$\Delta M_{q}$ in quadrature to find $\pm 6.8\%$ and $\pm 4.9\%$ for
(\ref{R3}) and (\ref{R4}), respectively. We have then added linearly 
the error of $\pm9\%$ from $\hat B_{B_q}$. Consequently the total 
uncertainties in $Br(B_{s}\to\mu\bar\mu)$ and $Br(B_{d}\to\mu\bar\mu)$ 
are found to be $\pm15.8\%$ and $\pm13.9\%$, respectively. If all errors
are added in quadrature we find $\pm11.3\%$ and $\pm10.2\%$, respectively.
As the errors in $\tau(B_{s})$, $\overline{m}_t(m_t)$ and 
$\Delta M_{s}$ will be decreased considerably in the coming years, the 
only significant errors in (\ref{R3}) and (\ref{R4}) will be then due to   
the uncertainties in $\hat B_{B_q}$. 
Future lattice calculations should be able 
to reduce these errors as well, so that predictions for 
$Br(B_{s,d}\to\mu\bar\mu)$ in the SM will become very accurate and 
in other MFV their accuracy will mainly depend on the knowledge of the
short distance functions $S$ and $Y$.

{\bf 4.} The dependence on new physics in (\ref{R2}) is given entirely by 
the ratio
$Y^2/S$. As hadronic uncertainties in (\ref{R2}) are substantially smaller
than in (\ref{DMQ}) and (\ref{bbll}), the differences between various MFV 
models can be easier seen. For instance in the ACD model with five dimensions 
\cite{appelquist:01} the ratio $Y^2/S$ with
$\overline{m}_t(m_t)=167\gev$ equals $0.58$, $0.53$, $0.49$ and $0.46$ for 
the compactifications scales $1/R=200,~250,~300,~400\gev$, respectively 
\cite{BSW02}. In the SM one has $Y^2/S=0.40$ and the effects of the 
Kaluza-Klein modes could in principle be seen when (\ref{R2}) is used, 
whereas it is very difficult by means of (\ref{bbll}).

The relation (\ref{R1}) that is satisfied in any model with MFV violation 
as defined in \cite{UUT},
is not satisfied in more 
complicated models in which other operators are relevant. As an example 
in the MSSM with MFV but large $\tan\beta$ the contributions of new LR scalar 
operators originating in neutral Higgs exchanges modify
$Br(B_{s,d}\to\mu\bar\mu)$ by orders of magnitude \cite{TT}, 
change $\Delta M_s$ typically 
by $10\%-30\%$ \cite{BUCHROSL}, 
leaving $\Delta M_d$ essentially unchanged with respect to 
the SM estimates. While a correlation between new physics effects in 
$Br(B_{s,d}\to\mu\bar\mu)$ and $\Delta M_s$ exists \cite{BUCHROSL}, 
the absence of
relevant new contributions to $\Delta M_d$ results in the violation of 
(\ref{R1}). Using the formulae of \cite{BUCHROSL} we find
\be\label{R5}
\frac{Br(B_{s}\to\mu\bar\mu)}{Br(B_{d}\to\mu\bar\mu)}
=\frac{\hat B_{d}}{\hat B_{s}}
\frac{\tau( B_{s})}{\tau( B_{d})} 
\frac{\Delta M_{s}}{\Delta M_{d}}  
\left[\frac{m_{B_s}}{m_{B_d}}\right]^4\frac{1}{1+f_s}
\ee
with $f_s$ being a complicated function of supersymmetric parameters 
that enters $\Delta M_s$ in this model. The dependence 
on $m_{B_q}$ in (\ref{R5}) originates in the LR scalar operators 
that dominate $Br(B_{s,d}\to\mu\bar\mu)$ at large $\tan\beta$.
Similarly, in a scenario considered in \cite{FIM03} in which new 
physics 
effects are assumed to be important in $\Delta M_d$ but negligible in 
$Br(B_{s,d}\to\mu\bar\mu)$ and $\Delta M_s$, the relation (\ref{R1})
is violated. Finally, we expect that it is generally violated in models with 
non-minimal flavour violation  \cite{Masiero,MIPORO,GRNIRA,DEPI}.

{\bf 5.} In summary we have presented stringent relations between
$Br(B_{s,d}\to\mu\bar\mu)$ and $\Delta M_{s,d}$ that are 
valid in the MFV models as defined in \cite{UUT}. The virtue of these 
relations is their
theoretical cleanness that allows to obtain improved predictions for 
$Br(B_{s,d}\to\mu\bar\mu)$ as demonstrated above. 
Other useful relations in the MFV models can be found in \cite{REL}.
 It will be interesting to  
follow the developments at Tevatron, LHC, BTeV, BaBar, Belle and 
K physics dedicated experiments to see whether these relations
are satisfied. While the present experimental upper bounds are still 
rather weak
\be
Br(B_{s}\to\mu\bar\mu) <2.6\times 10^{-6}~~~(95\%~{\rm C.L.}~~\cite{CDF}),
\ee
\be
Br(B_{d}\to\mu\bar\mu)<2.0\times 10^{-7}~~~(95\%~{\rm C.L.}~~\cite{BAB}),
\ee
considerable progress is expected in the coming years.

Needless to say the improvement on the accuracy of $F_{B_q}$ is very
important as $F_{B_s}/F_{B_d}$ is crucial for the determination of 
the CKM element $\vtd$ as seen in (\ref{dmsdmd}) and (\ref{bmumu}). 
Moreover the measurements of $\Delta M_s$ 
and $Br(B_{s}\to\mu\bar\mu)$ in
conjunction with  accurate values of  $\sqrt{\hat B_{s}}F_{B_s}$
and $F_{B_s}$ will determine the function $S$ and $Y$, respectively. This 
information will allow one to distinguish between various MFV models.

We would like to thank A. Dedes and  F. Kr{\"u}ger for useful 
remarks on the manuscript.
This
research was partially supported by the German `Bundesministerium f\"ur 
Bildung und Forschung' under contract 05HT1WOA3 and by the 
`Deutsche Forschungsgemeinschaft' (DFG) under contract Bu.706/1-2.

\renewcommand{\baselinestretch}{0.95}

\vfill\eject

\end{document}